\documentstyle[12pt]{article}

\newcommand{\sect}[1]{\setcounter{equation}{0}\section{#1}}

\def\bseq{\begin{subequation}}  
\def\eseq{\end{subequation}}
\def\bsea{\begin{subeqnarray}}  
\def\esea{\end{subeqnarray}}

\newcommand{\beq}{\begin{equation}}
\newcommand{\eeq}{\end{equation}}
\newcommand{\bea}{\begin{eqnarray}}
\newcommand{\eea}{\end{eqnarray}}
\newcommand {\non}{\nonumber}

\renewcommand{\a}{\alpha}
\renewcommand{\b}{\beta}

\renewcommand{\d}{\delta}
\newcommand{\th}{\theta}
\newcommand{\Th}{\Theta}

\newcommand{\di}{\partial}
\newcommand{\g}{\gamma}
\newcommand{\G}{\Gamma}

\newcommand{\e}{\epsilon}
\newcommand{\z}{\zeta}

\newcommand{\k}{\kappa}

\renewcommand{\l}{\lambda}

\newcommand{\p}{\pi}

\newcommand{\s}{\sigma}

\renewcommand{\o}{\omega}
\renewcommand{\O}{\Omega}

\newcommand{\calE}{{\cal E}}

\def\Mb{\kern 2pt\mathchoice
            {
             \vbox{\hrule width10pt height 0.4pt depth 0pt
                 \kern 1.2pt\hbox{\kern -2pt$\displaystyle M$}}}
            {
                 \vbox{\hrule width10pt height 0.4pt depth 0pt
                 \kern 1.2pt\hbox{\kern -2pt$\textstyle M$}}}
            {
\vbox{\hrule width6pt height 0.4pt depth 0pt
                 \kern 1.0pt\hbox{\kern -2pt$\scriptstyle M$}}}
            {
                 \vbox{\hrule width5pt height 0.4pt depth 0pt
                 \kern 0.8pt\hbox{\kern -2pt$\scriptscriptstyle M$}}}}

\def\Sb{\kern 2pt\mathchoice
            {
                 \vbox{\hrule width6pt height 0.4pt depth 0pt
                 \kern 1.2pt\hbox{\kern -2pt$\displaystyle S$}}}
            {
                 \vbox{\hrule width6pt height 0.4pt depth 0pt
                 \kern 1.2pt\hbox{\kern -2pt$\textstyle S$}}}
            {
                 \vbox{\hrule width3.5pt height 0.4pt depth 0pt
                 \kern 1.0pt\hbox{\kern -2pt$\scriptstyle S$}}}
            {
                 \vbox{\hrule width3pt height 0.4pt depth 0pt
                 \kern 0.8pt\hbox{\kern -2pt$\scriptscriptstyle S$}}}}

\def\Rb{\kern 2pt\mathchoice
            {
                 \vbox{\hrule width5.5pt height 0.4pt depth 0pt
                 \kern 1.2pt\hbox{\kern -2.5pt$\displaystyle R$}}}
            {
                 \vbox{\hrule width5.5pt height 0.4pt depth 0pt
                 \kern 1.2pt\hbox{\kern -2.5pt$\textstyle R$}}}
            {
                 \vbox{\hrule width3.5pt height 0.4pt depth 0pt
                 \kern 1.0pt\hbox{\kern -2.2pt$\scriptstyle R$}}}
            {
                 \vbox{\hrule width3pt height 0.4pt depth 0pt
                 \kern 0.8pt\hbox{\kern -2.2pt$\scriptscriptstyle R$}}}}

  \def\pp{{\mathchoice
          {
              \kern 1pt%
              \raise 1pt
              \vbox{\hrule width5pt height0.4pt depth0pt
                    \kern -2pt
                    \hbox{\kern 2.3pt
                          \vrule width0.4pt height6pt depth0pt
                          }
                    \kern -2pt
                    \hrule width5pt height0.4pt depth0pt}%
                    \kern 1pt
           }
            {
              \kern 1pt%
              \raise 1pt
              \vbox{\hrule width4.3pt height0.4pt depth0pt
                    \kern -1.8pt
                    \hbox{\kern 1.95pt
                          \vrule width0.4pt height5.4pt depth0pt
                          }
                    \kern -1.8pt
                    \hrule width4.3pt height0.4pt depth0pt}%
                    \kern 1pt
            }
            {
              \kern 0.5pt%
              \raise 1pt
              \vbox{\hrule width4.0pt height0.3pt depth0pt
                    \kern -1.9pt  
                    \hbox{\kern 1.85pt
                          \vrule width0.3pt height5.7pt depth0pt
                          }
                    \kern -1.9pt
                    \hrule width4.0pt height0.3pt depth0pt}%
                    \kern 0.5pt
            }
            {
              \kern 0.5pt%
              \raise 1pt
              \vbox{\hrule width3.6pt height0.3pt depth0pt
                    \kern -1.5pt
                    \hbox{\kern 1.65pt
                          \vrule width0.3pt height4.5pt depth0pt
                          }
                    \kern -1.5pt
                    \hrule width3.6pt height0.3pt depth0pt}%
                    \kern 0.5pt
            }
        }}

  \def\mm{{\mathchoice
                       {
                             \kern 1pt
               \raise 1pt    \vbox{\hrule width5pt height0.4pt depth0pt
                                  \kern 2pt
                                  \hrule width5pt height0.4pt depth0pt}
                             \kern 1pt}
                       {
                            \kern 1pt
               \raise 1pt \vbox{\hrule width4.3pt height0.4pt depth0pt
                                  \kern 1.8pt
                                  \hrule width4.3pt height0.4pt depth0pt}
                             \kern 1pt}
                       {
                            \kern 0.5pt
               \raise 1pt
                            \vbox{\hrule width4.0pt height0.3pt depth0pt
                                  \kern 1.9pt
                                  \hrule width4.0pt height0.3pt depth0pt}
                            \kern 1pt}
                       {
                           \kern 0.5pt
             \raise 1pt  \vbox{\hrule width3.6pt height0.3pt depth0pt
                                  \kern 1.5pt
                                  \hrule width3.6pt height0.3pt depth0pt}
                           \kern 0.5pt}
                       }}

\def\pd{{\kern0.5pt
                   + \kern-5.05pt \raise5.8pt\hbox{$\textstyle.$}\kern
0.5pt}}

\def\pmd{{\kern0.5pt
                  \pm \kern-5.05pt \raise6.3pt\hbox{$\textstyle.$}\kern1.5pt}}

\def\md{{\mathchoice
   {
      {{\kern 1pt - \kern-6.2pt \raise5pt\hbox{$\textstyle.$}\kern 1pt}}}
    {
      {{\kern 1pt - \kern-6.2pt \raise5pt\hbox{$\textstyle.$}\kern 1pt}}}
    {
      {\kern0.5pt - \kern-5.05pt \raise3.4pt\hbox{$\textstyle.$}\kern0.5pt}}
    {
      {\kern0.5pt - \kern-5.05pt \raise3.4pt\hbox{$\textstyle.$}\kern0.5pt}}}}

\newcommand{\Del}{\nabla}

\newcommand{\calD}{\cal D}

\def\Sc{\scriptstyle}

\newcommand{\reff}[1]{(\ref{#1})}

\newcommand{\shalf}{{\Sc\frac{1}{2}}}
\newcommand{\sihalf}{{\Sc\frac{i}{2}}}
\newcommand{\half}{\frac{1}{2}}
\newcommand{\ihalf}{\frac{i}{2}}
\newcommand{\squart}{{\Sc\frac{1}{4}}}

\renewcommand{\thefootnote}{\fnsymbol{footnote}}

\begin{document}

\newpage
\begin{titlepage}
\begin{flushright}
{WATPHYS-TH-97-07}\\
{hep-th/9708126}\\
\end{flushright}
\vspace{2cm}
\begin{center}
{\bf {\large} Supergravity from a Massive Superparticle and the Simplest
Super Black Hole}\\
\vspace{1.5cm}
Marcia
 E. Knutt-Wehlau\footnote{marcia@avatar.uwaterloo.ca.  John Charles Polanyi
Fellow.  Address after Sept. 1: Physics Dept., McGill University, Montreal, PQ
CANADA H3A 2T8}\\
\vspace{1mm}
and\\
\vspace{1mm}
R. B. Mann\footnote{mann@avatar.uwaterloo.ca} \\
\vspace{1mm}
{\em Physics Department, University of Waterloo, Waterloo, ON CANADA  N2L 3G1}

\vspace{1.1cm}
{{ABSTRACT}}
\end{center}

\begin{quote}

We describe in superspace a theory of a massive superparticle
coupled to a version of two dimensional $N=1$ dilaton supergravity.
The $(1+1)$ dimensional supergravity is generated by
the stress-energy of the superparticle, and the evolution of the superparticle
is reciprocally influenced by the supergravity.
We obtain exact superspace solutions for both the
superparticle worldline and the supergravity fields.
We use the resultant non-trivial compensator
superfield solution to construct a  model of a two-dimensional
supersymmetric black hole.
\end{quote}

\vfill

\begin{flushleft}
August 1997

\end{flushleft}
\end{titlepage}

\newpage

\renewcommand{\thefootnote}{\arabic{footnote}}
\setcounter{footnote}{0}
\newpage
\pagenumbering{arabic}

\sect{Introduction}

Superparticles \cite{Brink} have been useful in extracting some
fundamental features of superstring theory \cite{GSpart} as well as
some  basic properties of topological defects
of supersymmetric field theories \cite{Gaun}.
A $p$-dimensional defect is referred to as a $p$-brane, the
$p=0$ case being the superparticle. Recently there has been a resurgence
of interest in the study of superparticles in various dimensions, particularly
for the massive case, because of the connection to D-branes \cite{deWit}.

There is, however, a general dearth in the literature of exact non-trivial
solutions
(let alone superspace ones) to supersymmetric problems, even for classical
field theories. For classical supergravity, the solutions include wave-type
solutions
\cite{wave,RoseMat},
a supersymmetric generalization of an extreme Reissner-Nordstrom
black hole \cite{sblack}, and a point supercharge on a flat background
\cite{VolGalt}. These solutions are non-trivial in the sense that they cannot
be
reduced by supersymmetry transformations to purely bosonic solutions. However
they also  have either vanishing torsion or vanishing gravitino stress-energy
\cite{VolGalt}.

Motivated by the above, we consider in this paper the coupling of a massive
superparticle
to $N=1$ supergravity in $(1+1)$ dimensions. The comparative simplicity of 2D
$N=1$
supergravity makes it a natural theoretical laboratory for investigating
 how superparticles (and by extension, super $p$-branes) influence the
behaviour of supergravity fields and spacetime, and vice versa.
 In this case, superparticle-supergravity coupling
necessarily involves a dilatonic theory of supergravity, since the
Einstein-Hilbert
action is a topological invariant \cite{BanksMann}.  In general this would mean
that both the supergravity fields and the dilaton influence the evolution of
the
superparticle.  However, as mentioned below, there is a form
of dilatonic supergravity in which the dilaton does not
influence the classical evolution of the vielbein and superparticle.

This theory is a supersymmetric
generalization of the $(1+1)$ dimensional ``R=T'' theory, which has been of
particular interest insofar as it has a consistent Newtonian limit \cite{r3}
(a nontrivial issue for a generic dilaton gravity theory \cite{jchan}).
It also has an interesting set of solutions for many physical
situations
which closely resemble their $(3+1)$ dimensional counterparts
\cite{r4,tom,arnold,r5}.
The dilatonic part of the action is chosen so that the dilaton field
decouples from the classical field
equations. This ensures that the evolution of the
gravitational field is determined only
by the matter stress-energy (and reciprocally) \cite{r3,r4}, thereby
capturing the essence of general relativity (as opposed to
classical scalar-tensor theories) in two spacetime dimensions.
Indeed it is possible to interpret ``R=T'' theory as the $D\rightarrow 2$ limit
of
general relativity (as opposed to some particular solution(s)) \cite{2dross}.

In super ``R=T'' theory, the dilaton field classically decouples from the
evolution of the
supergravity/matter system.  Hence we have $(1+1)$ dimensional supergravity
being
generated by supersymmetric matter, and the evolution of the supermatter being
influenced by
the supergravity.  For the case in which the supermatter is given
by a single massive superparticle, we obtain  exact non-trivial
superspace solutions for
both the
superparticle worldline and the supergravity fields.

The outline of our paper is as follows. In section 2 we recapitulate the basic
 formalism for a massive superparticle in flat superspace, and in section 3 we
couple the superparticle to supergravity. In section 4 we describe
the form of dilaton ``R=T'' supergravity in superconformal gauge.
In section 5 we examine the superparticle action in superconformal gauge and
in sections 6 and 7 we solve for the supergravity compensator
 in the presence of a single superparticle. In section 8 we construct a model
of a
super black hole, based on the background generated by the compensator.
The appendices contain the equations of motion, some
component results, a discussion of the non-triviality of the solution,
 as well as the precise form of the super black hole vielbein.
We close with some concluding remarks.

\sect{Massive Superparticle In Flat Superspace}

  The standard covariant action for the massive superparticle in two dimensions
is
\beq
I = -m \int d\tau \Bigl[\sqrt{ -\p^n \p_n }
 + \ihalf \bar{\th}\Gamma^3 \dot{\th}\Bigr]
\eeq
with $\p^n = \dot{x}^n + \ihalf \bar{\th}\G^n \dot{\th}$, where
 $(\G^0)_{\a\b} = \Bigl( \begin{array}{cc} -1 & 0 \\
                           0 & -1 \\
          \end{array} \Bigr),
(\G^1)_{\a \b} =  \Bigl( \begin{array}{cc} -1 & 0 \\
                           0 & 1 \\
          \end{array} \Bigr)$ and $
(\G^3)_{\a \b} =  \Bigl( \begin{array}{cc} 0 & -1 \\
                           -1 & 0 \\
          \end{array} \Bigr)$.  The metric is $\eta_{ab} = (-, +)$, for $a = 0,
1$, and spinor
 indices
are raised and lowered by $\e_{\a\b} = \e^{\a\b} = \Bigl( \begin{array}{cc} 0 &
1 \\
                           -1 & 0 \\
          \end{array} \Bigr),$ where $\bar{\th}^\a = \th^\a = \e^{\a\b}
\th_\b.$
The action is manifestly globally supersymmetric, reparametrization invariant
and locally
kappa-invariant.

For convenience we switch to light-cone coordinates,
$(x^\pp, x^\mm)= \half (x^0 \pm x^1)$, by defining also
$(\p^\pp, \p^\mm) = \half(\p^0 \pm \p^1)$ and identifying
$\th^1 = \th^+$ and $\th^2 = \th^-$, so that in this notation the above action
becomes
\beq
I = -m \int d\tau \Bigl[\sqrt{\p^\pp \p^\mm} - \ihalf(\th^+ \dot{\th}^- + \th^-
\dot{\th}^+)\Bigr]
\eeq
where
$\p^\pp = \dot{x}^\pp + i \th^+ \dot{\th}^+, \p^\mm = \dot{x}^\mm + i \th^-
\dot{\th}^-.$
 We write the kappa-symmetry ($\k = \k(\tau)$) explicitly as
\bea
\d_\k x^\pp = - i\th^+ \delta_\kappa \th^+  &,& \d_\k x^\mm = - i\th^-
\delta_\kappa \th^-
\eea
and
\bea
\d_\k \th^+ = - \left( \k^+ + \k^- \frac{\p^\pp}{\sqrt{\p^2}} \right) &,&
\d_\k \th^- = - \left( \k^- + \k^+ \frac{\p^\mm}{\sqrt{\p^2}} \right)  ~~  .
\label{kappa}
\eea
  The equations of motion are:
\bea
\sqrt{\frac{\p^\mm}{\p^\pp}} = a ~~,~~ \sqrt{\frac{\p^\pp}{\p^\mm}} = b \non \\
 \dot{\th}^- = a \dot{\th}^+  ~~,~~ \dot{\th}^+ = b \dot{\th}^-
\eea
where $ a, b$ are constants and $ab = 1$.
{F}rom \reff{kappa}, it is possible to choose a gauge in which
one of the
$\th$'s is a constant, i.e., we take $\dot{\th}^- = 0$.  This implies that
$\dot{\th}^+ = 0$ likewise, so that both $\th^+$ and $\th^-$ are constants.
Note that setting
one of the thetas to zero is too strong a choice (and breaks manifest
supersymmetry) \cite{ulf}.
For a free particle we have $\p^\pp = c$ and $\p^\mm = d$,
both constants, so that $\pi^2 = cd = a^2c^2$ for a massive superparticle.

 To make contact with the standard Green-Schwarz superstring formulation, we
eliminate the square root from the action by introducing an einbein, $g$, on
the worldline of
the superparticle, and obtain the action in the usual form
\beq
I = -m \int d\tau \Bigl[g^{-1}\p^\pp \p^\mm - \ihalf(\th^+ \dot{\th}^- + \th^-
\dot{\th}^+) +
 \frac{g}{4} \Bigr]  ~~.
\eeq
  Varying with respect to $g$ gives $g = 2\sqrt{\p^\pp \p^\mm}$, and varying
with respect to the
particle coordinates gives back the previous equations of motion.   The
kappa-invariance is as
before with the variation of the einbein given by
\beq
\d_\k g = 4i(\dot{\th}^+ \k^- + \dot{\th}^- \k^+)   ~~.    \label{einkappa}
\eeq
   En route to coupling the superparticle to supergravity, we introduce
coordinates
 with world indices
$z^M = (x^m, \th^\mu)$, and the flat vielbein ${e_M}^A$, defined so
that $\dot{e}^A \equiv \dot{z}^M {e_M}^A = (\dot{x} + i \th \dot{\th},
\dot{\th})$.  We also
introduce a gauge field $\G_A = (\G_\a, \G_a)$ to describe  the Wess-Zumino
type term in the
action.  The action can then
be rewritten as
\bea
I &=& -m \int d\tau \Bigl[ g^{-1} \dot{z}^M {e_M}^\pp  \dot{z}^N {e_N}^\mm
\eta_{\pp \mm} +
 \dot{z}^M {e_M}^A \G_A + \frac{g}{4}  \Bigr] \non \\
&=& -m \int d\tau \Bigl[ g^{-1} \dot{e}^\pp \dot{e}^\mm + \dot{e}^A \G_A +
\frac{g}{4} \Bigr]
\eea
where $\G_+ = \ihalf \th^- , \G_- = \ihalf \th^+ , \G_a = 0$ and
$\eta_{\pp \mm} = 1$.

\sect{Coupling to Supergravity}

   We now couple the superparticle to supergravity. Promoting the
flat vielbein
to the curved one, ${{\calE}_M}^A$,  and $\G$ to a general superfield, the
action becomes
\bea
I_P &=& -m \int d\tau \left[ g^{-1} \dot{z}^M {{\calE}_M}^\pp  \dot{z}^N
{{\calE}_N}^\mm \eta_{\pp \mm} +
 \dot{z}^M {{\calE}_M}^A \G_A + \frac{g}{4}  \right] \non  \\
&=& -m \int d\tau \left[ g^{-1} \dot{\calE}^\pp \dot{\calE}^\mm +
\dot{\calE}^A \G_A + \frac{g}{4} \right]
\eea
We define the $\k$-transformations of the coordinates in curved superspace as
\bea
\d_\k {\calE}^a &\equiv& \d_\k z^M {{\calE}_M}^a = 0 \non \\
\d_\k {\calE}^\a &\equiv& \d_\k z^m {{\calE}_M}^\a
\eea
where explicitly we have
\bea\label{kappa1}
\d_\k {\calE}^+ = -  \bigl( \k^+ + 2 \frac{\k^-}{g} \dot{\calE}^\pp \bigr) &,&
\d_\k {\calE}^- = -  \bigl( \k^- + 2 \frac{\k^+}{g} \dot{\calE}^\mm \bigr) \non
\\
\eea
as well as
\bea\label{kappa2}
\d_\k g &=& 4i(\dot{\calE}^+\k^- + \dot{\calE}^-\k^+)  ~~.
\eea

  The supergravity covariant derivative is given by $\Del_A =
{{\calE}_A}^M \di_M
+ \O_A$, where $\O_A = \o_A M$ is the spin connection.
We use ordinary derivatives $\di_M$ for compatibility with the notation of
forms.
We define
 $\widehat{\Del}_A = \Del_A + \G_A$, including now the gauge field, and we have
\beq
[\widehat{\Del}_A, \widehat{\Del}_B\} = {T_{AB}}^C \widehat{\Del}_C + R_{AB} M
+ F_{AB}
\eeq
which defines the torsions, curvatures and gauge field strengths, respectively,
where $\di_{[M} {{\calE}_{N)}}^A = {T_{NM}}^A + {\o_{[MN)}}^A$.
Also,
\beq
F_{AB} = \Del_{[A} \G_{B)} - {T_{AB}}^C \G_C
\eeq
with $\{\G_A, \G_B ] = 0$. The constraints on $\G$ are
\bea
\G_\pp = - i \Del_+ \G_+  &,& \G_\mm = - i \Del_- \G_- \non \\
F_{+-}=F_{-+} &=& \Del_+ \G_- +\Del_- \G_+ = i  \label{Gammacons}
\eea
All other $F$'s are zero (as in the flat space
case),
consistent with the Bianchi identities \cite{jim}.

The variation of the action under a kappa transformation is given by
\beq
\d I_P = -m \int d\tau \Bigl\{\d g (-g^{-2}\dot{{\calE}^\pp} \dot{\calE}^\mm
+\frac{1}{4})
      + \dot{{\calE}^B} \d {\calE}^A [ g^{-1}(T_{BA}^\pp \dot{\calE}^\mm +
{T_{AB}}^\mm
 \dot{\calE}^\pp) +F_{AB}] \Bigr\}
\eeq
where we have used the following expressions in the derivation
\bea
\d \dot{\calE}^A &=& \d\dot{z}^M {{\calE}_M}^A + \dot{z}^M \d {{\calE}_M}^A
\non \\
             &=& \di_\tau (\d z^M {{\calE}_M}^A) - \d z^M \dot{z}^N \di_N
{{\calE}_M}^A +
\dot{z}^M \d z^N\di_N {{\calE}_M}^A \non\\
             &=& {\calD}_\tau(\d z^M {{\calE}_M}^A) + \d z^M \dot{z}^N T_{NM}^A
+\dot{\calE}^B \d z^M
 \o_{MB}^A
\eea
Substituting the explicit variations (\ref{kappa1}) and (\ref{kappa2}),
we find $\d I_P=0$, provided the supergravity constraints \reff{cons} and those
on $\G$,
\reff{Gammacons}, are
satisfied.

\sect{Dilaton Supergravity in Conformal Gauge}

We now couple the superparticle to $N=1$ dilaton supergravity, for which the
action is
\beq
I_D = \frac{1}{2\k}\int d^2xd^2 \th E^{-1}(\Del_+ \Phi \Del_- \Phi + \Phi R)
\\
\eeq
where $\Phi$ is the dilaton superfield,
 $R$ is the scalar supercurvature and $E = sdet {E_A}^M$.  We choose this
action as opposed to that given in \cite{park}, for example, since the dilaton
decouples from the evolution of the matter system in this case, and
gives the supersymmetric analogue of the bosonic ``R=T'' system \cite{r3, r4}.

The solution to the constraints is simplest in conformal gauge.  The
constraints are usually
solved in terms of covariant derivatives $\Del_A = {E_A}^M D_M + \o_A M$,
that are expanded with respect to the standard flat
supersymmetry covariant derivatives, $D_A = (D_+, D_-) = (\di_+ + i \th^+
\di_\pp,
\di_- + i \th^- \di_\mm)$.  However, the natural description for the
superparticle
is in terms of forms, and so we choose as a basis
the ordinary derivatives $\di_M = (\di_m, \di_\mu)$ as
mentioned earlier.  In this basis we write $\Del_A = {{\calE}_A}^M \di_M + \o_A
M$.
We solve the constraints
in conformal gauge in terms of the $D$'s and change
to the other basis afterwards.
 The (1,1) supergravity constraints \cite{jim, rocek} are
\bea
\{\Del_+, \Del_+\} = 2i \Del_\pp  &,& \{\Del_-, \Del_-\} = 2i \Del_\mm \non \\
\{\Del_+, \Del_-\} &=& RM \non \\
{T_{+\pp}}^A &=& {T_{-\mm}}^A = 0  \label{cons}
\eea
where the constraints on the covariant derivatives are solved in conformal
gauge
 in terms of the compensator
superfield $S$, as
\bea
\Del_+ = e^S [D_+ + 2 (D_+S)M] &,& \Del_- = e^S [D_- - 2 (D_-S)M] \non\\
\Del_\pp = e^{2S} [\di_\pp - 2i (D_+S)D_+ + 2 (\di_\pp S)M] &,&
\Del_\mm = e^{2S} [\di_\mm - 2i (D_-S)D_- - 2 (\di_\mm S)M] \non\\
R &=& 4e^{2S}D_- D_+ S \label{R}
\eea
{}From this we can read off the
elements of
${E_A}^M$ and compute $E^{-1} = e^{-2S}$.

  We now switch to the preferred basis and calculate the elements of
${{\calE}_A}^M$.
Inverting this matrix we obtain
\bea
{{\calE}_M}^A = \left[
\begin{array}{cccc}
e^{-2S} & 0 & 2ie^{-S}D_+S & 0 \\
0 & e^{-2S} & 0 & 2ie^{-S}D_-S \\
-ie^{-2S}\th^+ & 0 & e^{-S}(1-2(D_+S) \th^+) & 0 \\
0 & -i e^{-2S} \th^- & 0 & e^{-S}(1-2(D_-S) \th^-)
\end{array}
\right]
\eea
Therefore, in conformal gauge, the dilaton supergravity part of the action
becomes
\beq
I_D = \frac{1}{2\k}\int d^2x d^2\th (D_+ \Phi D_- \Phi + 4 \Phi D_- D_+ S)
\label{dil}
\eeq

 It is clear that the dilaton decouples from the supergravity-matter sector,
provided the matter is independent of the dilaton.  Indeed,
the equations of motion for the full action, $I=I_D(\Phi,S)+I_M(\Psi,S)$, where
$\Psi$ symbolically represents the supersymmetric matter sector, are
\bea
D_- D_+ \Phi + 2 D_- D_+ S  &=& 0\label{eqphi} \\
\frac{2}{\k} D_- D_+ \Phi +\frac{\delta I_M}{\delta S} &=& 0
\label{mateq}
\eea
along with the matter field equations of motion. The solution of \reff{eqphi}
 is $\Phi = -2S$, and inserting (\ref{eqphi})
into (\ref{mateq}) yields
\beq
-\frac{4}{\k} D_- D_+ S +\frac{\delta I_M}{\delta S} = 0
\label{mateq2}
\eeq
showing that the dilaton classically decouples from the supergravity-matter
system.

\sect{Superparticle Action in Conformal Gauge}

  The action is
\beq
I_P =  -m \int d^4z \int d\tau \left[ g^{-1} \dot{{z_0}}^M {{\calE}_M}^\pp
\dot{{z_0}}^N {{\calE}_N}^\mm
  + \dot{{z_0}}^M  {{\calE}_M}^A \G_A + \frac{g}{4} \right] \d (z -
{z_0}(\tau))
\eeq
where $z = (x, \th)$ are the coordinates of the superspace, and ${z_0}(\tau) =
( {x_0}(\tau),
{\th_0}(\tau))$ are the coordinates of the superparticle.
 We require the constraints on $\G$ in conformal gauge.
We define $\{\widehat{\Del}_+, \widehat{\Del}_+\} \equiv 2i
\widehat{\Del}_
\pp$, and similarly
for $\mm$,  which implies that
\bea
\G_\pp &=& - ie^S(D_+ \G_+ + (D_+S)\G_+)  \non \\
\G_\mm &=& - ie^S(D_- \G_- + (D_-S)\G_-) \label{vecGamcons}
\eea
Substituting in for $\calE$ and $\G$ we obtain
\bea
I_P &=&  -m \int d^4z \int d\tau \left\{ g^{-1} e^{-4S}(\dot{{x_0}}^\pp + i
{\th_0}^+ \dot{{\th_0}}^+)
            (\dot{{x_0}}^\mm + i {\th_0}^- \dot{{\th_0}}^-) \right.\non \\
            &+& i e^{-S}(\dot{{x_0}}^\pp + i {\th_0}^+ \dot{{\th_0}}^+)[(D_+S)
\G_+ - D_+ \G_+] \non \\
            &+& i e^{-S}(\dot{{x_0}}^\mm + i {\th_0}^- \dot{{\th_0}}^-)[(D_-S)
\G_- - D_- \G_-] \non\\
            &+&\left. e^{-S}(\dot{{\th_0}}^+ \G_+  + \dot{{\th_0}}^- \G_-)  +
\frac{g}{4}
       \right\}\d (z-{z_0}(\tau))
\label{part}
\eea
The complete action is the sum of \reff{dil} and \reff{part}.

  It is convenient to define $G_\a = e^S \G_\a$ and include \reff{Gammacons}
in the supergravity action by means of a lagrange multiplier, $\l$.
We obtain
\bea
I_P &=&  -m \int d^4z \int d\tau \left[ g^{-1} e^{-4S}(\dot{{x_0}}^\pp + i
{\th_0}^+ \dot{{\th_0}}^+)
            (\dot{{x_0}}^\mm + i {\th_0}^- \dot{{\th_0}}^-) \right. \non \\
            &+& i (\dot{{x_0}}^\pp + i {\th_0}^+ \dot{{\th_0}}^+)D_+G_+
                   + i (\dot{{x_0}}^\mm + i {\th_0}^- \dot{{\th_0}}^-)D_-G_-
\non\\
            &+& \left. \dot{{\th_0}}^+ G_+  + \dot{{\th_0}}^- G_-  +
\frac{g}{4} \right] \d (z-{z_0}(\tau))
 \label{act}
\eea
and
\bea
I_D = \frac{1}{2\k}\int d^2x d^2\th [D_+ \Phi D_- \Phi + 4 \Phi D_- D_+ S + \k
\l
e^{-2S}
        (D_+G_- + D_-G_+ - i e^{-2S})]  \label{condil}
\eea

  We now perform a change of variables in the superparticle action, by first
explicitly
writing it in terms of ${x_0}^0$ and ${x_0}^1$, and then making the gauge
choice ${x_0}^0 = \tau$ (static
gauge)
so that $\frac{\dot{{x_0}}^1}{\dot{{x_0}}^0} = \frac{d{x_0}^1}{d{x_0}^0} \equiv
\dot{{x_0}}$.  We also rename
$z^M = (x^0, x^1, \th^\mu) = (t, x, \th^\mu)$, so \reff{act} becomes
\bea
I_P &=&  -m \int dt dx d^2 \th \int d{x_0}^0 \left\{ g^{-1} e^{-4S}
\left[\shalf
(1+\dot{{x_0}}) +
            i {\th_0}^+ \dot{{\th_0}}^+ \right]
             \left[\shalf ( 1 -\dot{{x_0}}) + i {\th_0}^- \dot{{\th_0}}^-
\right] \right. \non \\
            &+& i \left[ \shalf ( 1 +\dot{{x_0}}) + i {\th_0}^+ \dot{{\th_0}}^+
\right]D_+G_+
                + i \left[ \shalf ( 1 -\dot{{x_0}}) + i {\th_0}^-
\dot{{\th_0}}^-\right] D_-G_-  \non\\
            &+& \left. \dot{{\th_0}}^+ G_+  + \dot{{\th_0}}^- G_-  +
\frac{g}{4} \right\}
            \d (t-{x_0}^0) \d (x-{x_0}({x_0}^0)) \d (\th^+ -
{\th_0}^+({x_0}^0)) \d (\th^- - {\th_0}^-({x_0}^0))\non \\
\eea
and
doing the ${x_0}^0$ integration gives
\bea
I_P &=&  -m \int dt dx d^2 \th \left\{ g^{-1} e^{-4S} \left[\shalf
(1+\dot{{x_0}}) + i {\th_0}^+
            \dot{{\th_0}}^+ \right]
            \left[\shalf ( 1 -\dot{{x_0}}) + i {\th_0}^- \dot{{\th_0}}^-\right]
\right. \non \\
            &+& i \left[ \shalf ( 1 +\dot{{x_0}}) + i {\th_0}^+ \dot{{\th_0}}^+
\right] D_+G_+
                 + i \left[ \shalf ( 1 -\dot{{x_0}}) + i {\th_0}^-
\dot{{\th_0}}^- \right] D_-G_-  \non\\
            &+& \left. \dot{{\th_0}}^+ G_+  + \dot{{\th_0}}^- G_- + \frac{g}{4}
\right\} \d (x-{x_0}(t))
            \d (\th^+ - {\th_0}^+(t)) \d (\th^- - {\th_0}^-(t))
\eea

  The equations of motion are given in detail in Appendix A, and we just
mention the
general content here.  Varying with respect to $\Phi$ shows that $\Phi$
decouples from the
action; varying with respect to $\lambda$ gives back the constraint on $G$; and
varying
 with respect to
$g$ allows the elimination of the einbein from the action.  The main equation
of motion
is the one for $S$, \reff{S}.

  In section 2, we listed the equations of motion for the free
superparticle in
flat superspace.  We reconsider those equations, written now in terms
of the new
variables.  We have
\bea
\dot{{x_0}}^\pp + i {\th_0}^+ \dot{{\th_0}}^+ &=& \shalf(\dot{{x_0}}^0 +
\dot{{x_0}}^1) + i {\th_0}^+ \dot{{\th_0}}^+ = c \non\\
\dot{{x_0}}^\mm + i {\th_0}^- \dot{{\th_0}}^- &=& \shalf(\dot{{x_0}}^0 -
\dot{{x_0}}^1) + i {\th_0}^- \dot{{\th_0}}^- = d \non\\
\eea
which become
\bea
 \shalf(1 + \dot{{x_0}}) + i {\th_0}^+ \dot{{\th_0}}^+ &=& c \non \\
 \shalf(1 - \dot{{x_0}}) + i {\th_0}^- \dot{{\th_0}}^- &=& d
\eea
When $\dot{{\th_0}} = 0$, the free superparticle moves  with a constant
velocity
$\dot{{x_0}} = 2c-1= 1-2d.$

\sect{Solution for Compensator and Motion of Superparticle}

In solving this problem classically, we look for an explicit expression
for the compensator $S$ in the dilaton supergravity, that is consistent with
the motion of the superparticle. We solve \reff{S} by analogy with the
bosonic case.  The observation that is crucial
in the solution of the
equations \reff{phi} to \reff{Gm} is that
 if the solution for $S$ and its derivatives vanishes on the superparticle's
worldline,
 then the equations
of motion reduce to those for a free superparticle in flat superspace.
We look for such a solution.

  Using this and noting that on the worldline, where $S=0$,  the constraint
on the $G$'s becomes trivial, we obtain
\bea
& D_- D_+ S(z) & \non \\
   &=& \frac{\k m}{2} \int dt \left\{ g^{-1} \left[\shalf
(1+\dot{{x_0}})
         + i {\th_0}^+ \dot{{\th_0}}^+ \right]
            \left[ \shalf ( 1 -\dot{{x_0}}) + i {\th_0}^-
\dot{{\th_0}}^-\right] \right\} \d^4 (z - {z_0}(t))
      \non\\
&=& \frac{\k m}{2} \sqrt{\pi^2} \d (x-{x_0}(t)) \d (\th^+ - {\th_0}^+(t))
\d(\th^- - {\th_0}^-(t))
\label{comp}
\eea
where $ \sqrt{\pi^2} = \shalf \sqrt{1-\dot{{x_0}}^2}$ for a free particle.

 We consider first the case of a superparticle at rest and assume that
 $S$ is time independent. Expanding it in a
power
series in $\th$, we have
\bea
S(x, \th) &=& -\frac{{\k}m}{4} [f(x) + \th^+ g(x) + \th^- h(x) + \th^+ \th^-
k(x)] \label{expan}
\eea
Substituting \reff{expan} into \reff{comp}
 we find the following differential equations for the component fields
\bea
f'' &=& - \d(x-x_0) \non \\
g' &=& i {\th_0}^+ \d(x-x_0) \non \\
h' &=& -i {\th_0}^- \d(x-x_0) \non \\
k &=& {\th_0}^+ {\th_0}^- \d(x-x_0)
\eea
which are solved by
\bea
f &=& -\shalf |x-{x_0}| =  -\shalf [\Th (x-{x_0})(x-{x_0}) + \Th
({x_0}-x)({x_0}-x)] \non\\
g &=& \sihalf {\th_0}^+[\Th(x-{x_0}) -\Th({x_0}-x)] \non \\
h &=& -\sihalf {\th_0}^-[\Th(x-{x_0}) -\Th({x_0}-x)] \non \\
k &=& {\th_0}^+ {\th_0}^- \d (x-{x_0})
\eea
where $\Th$ is the Heaviside function.
Therefore we can write $S$ as
\bea
S &=&  \frac{{\k}m}{4}\{\shalf |x-{x_0}| + \sihalf (\th^+ {\th_0}^+ - \th^-
{\th_0}^-)[\Th
(x-{x_0}) -\Th ({x_0}-x)]
     + \th^+ \th^- {\th_0}^+ {\th_0}^- \d (x-{x_0})\} \non \\
&=& -\frac{{\k}m}{8}\{ |x-{x_0} - i(\th^+ {\th_0}^+ - \th^- {\th_0}^-)|
\}\label{ans}
\eea
where the second line is to be  understood as a Taylor expansion.
  Note that $S$ and its derivatives vanish on the worldline of the
superparticle,
 that is when
$x={x_0}$ and $\th= {\th_0}$.

 The solution for a moving superparticle can be obtained by a
 Lorentz boost of $x$, $t$ and $\th$, where we have
\bea
x' = \frac{x - \dot{{x_0}}t}{\sqrt{1 -\dot{{x_0}}^2}}  &,& t' = \frac{t -
\dot{{x_0}}x}{\sqrt{1 -\dot{{x_0}}^2}}
       \non \\
{x^\pp}' = \z^2 x^\pp &,& {x^\mm}' = \z^{-2} x^\mm  \non \\
{\th^+}' = \z \th^+ &,& {\th^-}' = \z^{-1} \th^-
\eea
with $\z^2 = \sqrt{\frac{1-\dot{{x_0}}}{1 + \dot{{x_0}}}}$.
Applying this to \reff{ans} we obtain a new Lorentz transformed $S'$
\bea
S' &=& -\frac{{\k}m}{8} \frac{1}{\sqrt{1 -\dot{{x_0}}^2}} |x-{x_0} - i(1 -
\dot{{x_0}})\th^+ {\th_0}^+
          + i (1 + \dot{{x_0}}) \th^- {\th_0}^-|
\eea
which gives a solution in which $\dot{{x_0}}=constant$
 and $\dot{{\th_0}} = 0$.
 We have as the final solution for the compensator
\bea
S &=& - \frac{\k m}{8} \frac{1}{\sqrt{1 -\dot{{x_0}}^2}} |x-{x_0}(t) - i(1 -
\dot{{x_0}})\th^+ {\th_0}^+(t)
          + i (1 + \dot{{x_0}}) \th^- {\th_0}^-(t))| \label{result}
\eea
It is straightforward to verify that \reff{result} gives the full solution to
\reff{comp}.

  We stress that this solution is non-trivial in that it cannot be obtained by
an
infinitesimal supersymmetry
transformation from the bosonic solution for a particle moving in dilaton
gravity.  We discuss this issue in Appendix C,  from both a superspace
 and a component viewpoint.

\sect{The Simplest Super Black Hole}

The solution obtained in the preceding section is the supersymmetric version of
that obtained for the gravitational field generated by a point particle in
$(1+1)$ dimensional R=T theory \cite{r3,tarasov}.  In conformal gauge the
metric
for this latter solution is \cite{tarasov}
\beq
ds^2 = e^{2m|z|}(-dt^2+dz^2)
\label{2dpoint}
\eeq
where the parameter $m$ is the mass of the particle. This solution can be
rewritten as
\beq
ds^2 = - (2m|w| + C)dt^2+ \frac{dw^2}{2m|w|+C} \label{2dschwzp}
\eeq
under a straightforward transformation of coordinates, where $C=1$.

Despite the fact that spacetime is flat everywhere outside of the particle,
we can use it to construct a two dimensional black hole using the methods
described in \cite{r3,tarasov}. Since
the Ricci scalar is $r=-4m \delta(w)$, independent of the sign of $C$,
the metric
\beq
ds^2 = - (2m|w| - |C|)dt^2+ \frac{dw^2}{2m|w|- |C|} \label{2dschwz}
\eeq
is also a solution of the field equation.  This is
a black hole whose event horizons are located at $w=\frac{|C|}{2m}$. This
black hole spacetime may be constructed by taking two copies of Minkowski
space, cutting each of them along the hyperboloids $T^2-X^2 = m^2$, and
gluing the spacetimes along their hyperboloids in a manner that does not
generate closed timelike curves ({\it i.e.} by gluing the hyperboloids
at positive $T$ together and the hyperboloids at negative $T$ together).
Details
of this construction are provided in refs. \cite{r4,tom}. A description
of how this black hole can be understood to arise as the endpoint of
gravitationally collapsing matter in $(1+1)$ dimensions is given in refs.
\cite{arnold,vendrell}.

We seek here the supersymmetric analogue of the spacetime described by
(\ref{2dschwz}), {\it i.e.} of a spacetime whose zweibein is
\bea\label{2dzweibos}
{e_m}^a = \left[
\begin{array}{cc}
\sqrt{\a} & 0  \\
0 & {\sqrt{\a}}^{-1}
\end{array}
\right]
\eea
where $\alpha \equiv 2m|w| + C$. For positive $C$ the zweibein is
that of a naked point particle of mass $m$, whereas for $C<0$ the zweibein is
that of a black hole.

The first step,
having found $S$, is to determine the vielbein in other than
superconformal coordinates - in particular, in superspace coordinates that are
the
analogue of those used in Schwarzschild gauge in the bosonic case
(\ref{2dschwz}).
We construct a model of a supersymmetric black hole by
examining the supercoordinate transformations that correspond
to the  ordinary $x$-space ones used to
construct bosonic two-dimensional black holes in ref. \cite{tarasov}.
Under a supercoordinate transformation $z=(x,\th) \rightarrow w=(u, \eta)$,
the vielbein transforms as ${\calE'_M}^A(w) = \frac{dz^N}{dw^M}
{\calE_N}^A(z)$, and we
demand that
in the transformed coordinates, the bosonic-bosonic corner of the
vielbein matrix (corresponding to ${\calE_m}^a$) have the same form
that the component zweibein ${e_m}^a$ does in the purely bosonic case
(\ref{2dzweibos}).

  We define the transformations by analogy with the bosonic case,
and we require $(x=x_0, \th=\th_0)$
to correspond to $(u=u_0, \eta=\eta_0)$ and also, $u>0 (u<0)$ when $x>0 (x<0)$.
This ensures that $\Th(x-x_0) - \Th(x_0-x) \equiv \e(x-x_0) = \e(u-u_0)$.
We expand each
side of the supercoordinate transformation equation in powers of either
 $\eta$ or $\th$, and determine the transformations
of the $\th$'s by matching both sides.

Consider first the case in which $C>0$.
We find that the supercoordinate transformation that relates the conformal
coordinates to the Schwarzschild coordinates is:
\beq
2m |u-u_0 - i(\eta^+ \eta_0^+ - \eta^- \eta_0^-)| + |C| = |C| e^{2m|x-x_0 -
       i(\th^+\th_0^+ -\th^- \th_0^-)|}
\eeq
which implies that
\bea
|x -x_0| &=& \frac{1}{2m} ln(\frac{2m}{|C|}|u-u_0| + 1) \non \\
\th^\pm &=& \frac{\sqrt{|C|}}{2m}\frac{\eta^\pm}{|u-u_0| + |C|/2m}\left[1 +
\sihalf(\eta^+ \eta_0^+
      - \eta^- \eta_0^-)\frac{\e(u-u_0)}{|u-u_0| + |C|/2m}\right] \non\\
\th_0^\pm &=& \frac{\eta^\pm_0}{\sqrt{|C|}} \non
\eea
 This case corresponds to just a massive superparticle at rest, and no black
hole.  We find
the scalar supercurvature in the new coordinate system to be
\beq
 R' = 8m \d(u-u_0) \d^{(2)}(\eta - \eta_0)
\eeq
and, as discussed in Appendix C, the  transformed $x$-space component curvature
is
\beq\label{ppcurvs}
r' = (\Del^2 R)'| = -8m|C| \d(u-u_0)
\eeq

For $C<0$, we find that the supercoordinate transformations that model a super
black
hole can be split up into three regions:

{\bf Region (i)}: $u-u_0 <\frac{-|C|}{2m}$
\beq
-2m[u-u_0 - i(\eta^+ \eta_0^+ - \eta^- \eta_0^-)] - |C| = |C| e^{-2m[x-x_0 -
i(\th^+\th_0^+ -\th^-
                              \th_0^-)]}
\eeq
which implies that
\bea
x-x_0 &=& \frac{-1}{2m} ln(\frac{-2m}{|C|}(u-u_0) - 1) \non \\
\th^\pm &=& \frac{i\sqrt{|C|}}{2m}\frac{\eta^\pm}{(u-u_0) + |C|/2m}\left[1 +
\sihalf(\eta^+ \eta_0^+
      - \eta^- \eta_0^-)\frac{1}{(u-u_0) + |C|/2m}\right] \non\\
\th_0^\pm &=& \frac{i\eta^\pm_0}{\sqrt{|C|}} \non
\eea

{\bf Region (ii)}: $|u-u_0| <\frac{|C|}{2m}$
\beq
2m |u-u_0 - i(\eta^+ \eta_0^+ - \eta^- \eta_0^-)| - |C| = - |C| e^{-2m|x-x_0
          - i(\th^+\th_0^+ -\th^- \th_0^-)|}
\eeq
which implies that
\bea
x -x_0 &=& \frac{-1}{2m} ln(1 - \frac{2m}{|C|}|u-u_0|) \non \\
\th^\pm &=& \frac{\sqrt{|C|}}{2m}\frac{\eta^\pm}{|u-u_0| - |C|/2m}\left[1 +
\sihalf(\eta^+ \eta_0^+
      - \eta^- \eta_0^-)\frac{\e(u-u_0)}{|u-u_0| - |C|/2m}\right] \non\\
\th_0^\pm &=& \frac{-\eta^\pm_0}{\sqrt{|C|}} \non
\eea

{\bf Region (iii)}: $u-u_0 >\frac{|C|}{2m}$
\beq
2m[u-u_0 - i(\eta^+ \eta_0^+ - \eta^- \eta_0^-)] - |C| = |C| e^{2m[x-x_0 -
i(\th^+\th_0^+ -\th^-
                              \th_0^-)]}
\eeq
which implies that
\bea
x-x_0 &=& \frac{1}{2m} ln(\frac{2m}{|C|}(u-u_0) - 1) \non \\
\th^\pm &=& \frac{i\sqrt{|C|}}{2m}\frac{\eta^\pm}{(u-u_0) - |C|/2m}\left[1 +
\sihalf(\eta^+ \eta_0^+
      - \eta^- \eta_0^-)\frac{1}{(u-u_0) - |C|/2m}\right] \non\\
\th_0^\pm &=& \frac{-i\eta^\pm_0}{\sqrt{|C|}}\non
\eea

  We can compute $R'$ and $r'$ for these cases and  we find
\bea\label{bhcurvs}
R' &=&  8m \d(u-u_0) \d^{(2)}(\eta - \eta_0) \non \\
r' &=& (\Del^2 R)'| = -8m|C| \d(u-u_0)
\eea
in region (ii), and $R'=r'=0$ in regions (i) and (iii).

Equations (\ref{ppcurvs}) and (\ref{bhcurvs}) show that the
supergravity solution we have obtained satisfies the field
equations independently  of the sign of $C$.  For $C<0$ we can
perform the same construction as in ref. \cite{r3,tom}, only
now in superspace. One takes two copies of 2D flat superspace,
cuts off the parts defined by spacelike bosonic hyperbolae,
and then glues them together so that there are no closed timelike
curves.  This yields the solution given by regions (i)--(iii) above.
Region (ii)  corresponds to the region inside the super black hole,
whereas regions
(i) and (iii) correspond to the region outside the super black hole.
The precise form of the
vielbein corresponding to each  region is given in
Appendix D.

We close this section by commenting on the $C=0$ case. For bosonic R=T theory
the analogous construction would involve gluing two copies of $|T|<|X|$
Minkowski spacetime along their $|T|=|X|$ lightcones.  The resultant manifold
would be non-Hausdorff at $X=T=0$. A possible resolution of this dilemma would
be
to glue only the right-hand Rindler wedges of each spacetime along their
respective
light cones, but it would be unclear how to avoid closed timelike lines in
a manner that yielded a consistent gluing at $T=0$.   We shall not consider
this
case any further.

\sect{Conclusions}

We have examined a $(1+1)$ dimensional dilaton supergravity theory
in which the dilaton classically decouples from the supergravity-matter system.
The stress-energy of the supermatter generates the supergravity, which in turn
governs the evolution of all supermatter fields.  We have found a non-trivial
solution for
the compensator superfield that describes the supergravity generated by a
massive
 superparticle. We have also shown how to construct
a two-dimensional super black hole from this supergravity solution.

A number of interesting questions arise from this work.  Apart from
being of interest in their own right, exact superspace solutions
might permit us to make considerable headway in the interpretation of classical
supergravity
solutions. These interpretative problems have received only scant attention
in the literature \cite{RoseMat,sblack} to date.

Recent progess in the $n$-body problem in $(1+1)$ dimensions \cite{Ohta}
suggests the possibility of making progress in solving the super $n$-body
problem in
two dimensions as well.  As demonstrated in ref.
\cite{Ohta},
the motion in the bosonic case is quite complicated even for $n=2$, and
contains
a variety of interesting features.

As a final comment, it would be of considerable interest to investigate
the influence of quantum corrections on the gravitational
and/or matter fields of such  (classical) solutions.

\noindent {\bf Acknowledgments}

  We would like to thank Jim Gates and Marc Grisaru for discussions. This
research was
supported in part by a John Charles Polanyi Fellowship, NSERC of Canada,
and an NSERC Postdoctoral Fellowship.

\vspace{0.5cm}
\appendix{\Large {\bf Appendices}}
\section{{\bf Equations of Motion}}
\setcounter{equation}{0}
 Varying the complete action  with respect to $\Phi, S, {x_0}, \th_0, G, \l$
and $g$ gives:
\beq
\Phi = -2S \label{phi}
\eeq
\bea
&&\frac{2}{\k} D_- D_+ \Phi(z) - \l e^{-2S}(D_+G_- + D_- G_+ - i e^{-2S}) \non
\\
         &+& 4m \int dt \left[ g^{-1} e^{-4S}[\half (1+\dot{{x_0}}) + i
{\th_0}^+ \dot{{\th_0}}^+]
            [\half ( 1 -\dot{{x_0}}) + i {\th_0}^- \dot{{\th_0}}^-] \right]
\d^4 (z - {z_0}(t)) = 0
\label{S} \non \\
\eea
\bea
&& \left\{ -4g^{-1} e^{-4S} \frac{\di S}{\di {x_0}} \left[ \shalf
(1+\dot{{x_0}}) + i {\th_0}^+
 \dot{{\th_0}}^+
              \right]
           \left[ \shalf ( 1 -\dot{{x_0}}) + i {\th_0}^- \dot{{\th_0}}^-
\right] \right. \non\\
 &+& i \left[\shalf (1+\dot{{x_0}}) + i {\th_0}^+ \dot{{\th_0}}^+
\right]\frac{\di (D_+G_+)}{\di {x_0}}
   + i \left[\shalf (1-\dot{{x_0}}) + i {\th_0}^- \dot{{\th_0}}^-
\right]\frac{\di (D_-G_-)}{\di {x_0}}\non \\
 &+& \dot{{\th_0}}^+ G_+  + \dot{{\th_0}}^- G_- \non \\
 &-& \left. \half \frac{d}{dt} \left[ g^{-1} e^{-4S}(-\dot{{x_0}} + i {\th_0}^-
\dot{{\th_0}}^-
       - i {\th_0}^+ \dot{{\th_0}}^+)
              + i D_+ G_+ - i D_- G_- \right] \right\} \d (t-s) = 0 \non
\label{{x_0}} \\
\eea
\bea
&& \left\{-4g^{-1} e^{-4S} \frac{\di S}{\di {\th_0}^+} \left[ \shalf
(1+\dot{{x_0}}) + i {\th_0}^+
    \dot{{\th_0}}^+\right] \left[\shalf ( 1 -\dot{{x_0}}) + i {\th_0}^-
\dot{{\th_0}}^- \right]  \right.\non\\
&+& g^{-1} e^{-4S}i \dot{{\th_0}}^+ \left[\shalf ( 1 -\dot{{x_0}}) + i
{\th_0}^- \dot{{\th_0}}^- \right]
    -\dot{{\th_0}}^+D_+ G_+  \non \\
&+& i \left[\shalf ( 1 -\dot{{x_0}}) + i {\th_0}^- \dot{{\th_0}}^- \right]
\frac{\di(D_-G_-)}{\di {\th_0}^+}
  + i \left[\shalf ( 1 +\dot{{x_0}}) + i {\th_0}^+ \dot{{\th_0}}^+ \right]
\frac{\di(D_+G_+)}{\di {\th_0}^+}\non\\
&-& \dot{{\th_0}}^+ \frac{\di G_+}{\di {\th_0}^+} - \dot{{\th_0}}^- \frac{\di
G_-}{\di {\th_0}^+} \non \\
&-& \left. \frac{d}{dt} \left[ g^{-1} e^{-4S}(\sihalf \dot{{\th_0}}^+ (1
-\dot{{x_0}})
    -\dot{{\th_0}}^+ {\th_0}^- \dot{{\th_0}}^-) + {\th_0}^+ D_+ G_+ + G_+
\right] \right\} \d (t-s) =0
 \label{thetap} \non \\
\eea
\bea
&& \left\{ -4g^{-1} e^{-4S} \frac{\di S}{\di {\th_0}^-} \left[\shalf
(1+\dot{{x_0}}) + i {\th_0}^+ \dot{{\th_0}}^+
      \right] \left[ \shalf ( 1 -\dot{{x_0}}) + i {\th_0}^- \dot{{\th_0}}^-
\right] \right. \non\\
&+& g^{-1} e^{-4S}i \dot{{\th_0}}^- \left[ \shalf ( 1 + \dot{{x_0}}) + i
{\th_0}^- \dot{{\th_0}}^- \right]
     -\dot{{\th_0}}^-D_- G_- \non \\
&+& i \left[ \shalf ( 1 -\dot{{x_0}}) + i {\th_0}^- \dot{{\th_0}}^-
\right]\frac{\di(D_-G_-)}{\di {\th_0}^-}
  + i \left[ \shalf ( 1 +\dot{{x_0}}) + i {\th_0}^+ \dot{{\th_0}}^+
\right]\frac{\di(D_+G_+)}{\di {\th_0}^-}\non\\
&-& \dot{{\th_0}}^+ \frac{\di G_+}{\di {\th_0}^+} - \dot{{\th_0}}^- \frac{\di
G_-}{\di {\th_0}^+} \non \\
&-& \left. \frac{d}{dt} \left[ g^{-1} e^{-4S}(\sihalf \dot{{\th_0}}^- (1
+\dot{{x_0}}) -\dot{{\th_0}}^+
   {\th_0}^+ \dot{{\th_0}}^-) + {\th_0}^- D_- G_- + G_- \right] \right\} \d
(t-s) =0 \label{thetam} \non\\
\eea
\bea
m \int dt \left[ i (\shalf ( 1 +\dot{{x_0}}) + i {\th_0}^+ \dot{{\th_0}}^+)D_+
\d^4 (z - {z_0}(t)) + \dot{{\th_0}}^+
  \d^4 (z - {z_0}(t)) \right] &=& \shalf D_-( \l e^{-2S})  \label{Gp} \non \\
m \int dt \left[ i (\shalf ( 1 -\dot{{x_0}}) + i {\th_0}^-\dot{{\th_0}}^-)D_-
\d^4 (z - {z_0}(t)) + \dot{{\th_0}}^-
  \d^4 (z - {z_0}(t)) \right] &=& \shalf D_+( \l e^{-2S}) \label{Gm} \non \\
\eea
\beq
 D_+G_- + D_-G_+ - i e^{-2S} = 0
\eeq
\bea
g &=& 2 e^{-2S}(\shalf ( 1 +\dot{{x_0}}) + i {\th_0}^+ \dot{{\th_0}}^+)
   (\half ( 1 -\dot{{x_0}}) + i {\th_0}^- \dot{{\th_0}}^-) \non \\
& = & 2 e^{-2S} \sqrt{\pi^\pp} \sqrt{\pi^\mm}
\eea
as the equations of motion for the respective fields.

\sect{Component Action}

  We obtain the component superparticle action by following the method used in
\cite{atick} for normal coordinate expansions of Green-Schwarz type
$\s$-models.
We modify the procedure slightly to bring the WZ-gauge choice in line with the
superconformal gauge choice, according to the discussion in the beginning of
Appendix C.
Following the notation of \cite{atick}, superspace is parametrized by $z^M =
(x_0^m, 0)$
 and $y^M = (0, y^\mu)$ and we replace the WZ-gauge choice,
 ${{\calE}_\mu}^\a =\d_\mu^\a$, with ${{\cal{E}}_\mu}^\a = e^{-S}\d_\mu^\a$.
Then
$y^\a = y^\mu{{\cal{E}}_\mu}^\a = {\th_0}^\a e^{-S}|$ is the only change to the
results
in the above references. Defining $V^A \equiv \dot{z}^M{{\cal{E}}_M}^A$,
 we find for the terms in the component superparticle action
\bea
I^{(0)} &=& -m \int d\tau [ g^{-1}V^\pp V^\mm + V^A \G_A + \frac{g}{4}] \non \\
I^{(1)} &=& -m \int d\tau [ -2ig^{-1}(V^\mm V^+ y^+ + V^\pp V^- y^-) + iV^+ y^-
                 +i V^-y^+]  \non \\
I^{(2)} &=& -m \int d\tau [ -2ig^{-1}(V^\mm {\cal{D}}y^+ y^+ + V^\pp
{\cal{D}}y^- y^-)
                +i({\cal{D}}y^+ y^- + {\cal{D}}y^- y^+)\non \\
         &~&  -2g^{-1}R V^\pp V^\mm y^+ y^- + 8 g^{-1} V^+ V^- y^+ y^-] \non \\
I^{(3)} &=& -m \int d\tau [ - 8 g^{-1}(V^- {\cal{D}}y^+y^+ y^- + V^+
{\cal{D}}y^- y^- y^+)]\non \\
I^{(4)} &=& -m \int d\tau [ -16g^{-1} {\cal{D}}y^+ y^+ {\cal{D}}y^-
y^+]\label{I}
\eea
where all quantities are evaluated at $(x_0^m, 0)$ and ${\cal{D}}$ is the
world-line
covariant derivative.

  Specifically for the superparticle, we have:
\bea
V^\pp &=& \shalf(1+\dot{x}_0) e^{-2S}| \non \\
V^\mm &=& \shalf(1-\dot{x}_0) e^{-2S}| \non \\
V^+ &=& -i(1+\dot{x}_0) D_+(e^{-S})|   \non \\
V^- &=& -i(1-\dot{x}_0) D_-(e^{-S})|   \non \\
{\cal{D}}y^\pm &=& ({\cal{D}} \th_0^\pm) e^{-S}| + \th_0^\pm ({\cal{D}}e^{-S})|
\eea
where $|=|_{\th_0 = 0}$.  We have also
${E_\mu} = e^{-S} D_\mu$, and we introduce ${\tilde{\psi}}$ with (lower world,
upper tangent)
indices, related to $\psi$ by multiplication by the component zweibein.  By
analogy with Appendix C
 we can therefore write
\bea
E_\pm &=& e^{-S} D_\pm \non \\
 &=& [ e^{1/4} + \sihalf(\th^+ {{\tilde{\psi}}_\pp}^+ - \th^-
{{\tilde{\psi}}_\mm}^-) +
\squart \th^+ \th^-
            e^{-1/4}(i\tilde{A} +  {{\tilde{\psi}}_\mm}^-
{{\tilde{\psi}}_\pp}^+) ] D_\pm
\eea
Therefore we have
\bea
D_+e^{-S}| &=& \sihalf {{\tilde{\psi}}_\pp}^+ \non \\
D_-e^{-S}| &=& -\sihalf {{\tilde{\psi}}_\mm}^- \non \\
e^{-S}| &=& e^{1/4} \non \\
y^\pm &=& \th_0^\pm e^{1/4} \non \\
\G_A| &=& \g_A
\eea
and substituting these expressions into \reff{I}, we obtain
\bea
I^{(0)} &=& -m \int d\tau [ \frac{g^{-1}}{4}(1+\dot{x}_0)(1-\dot{x}_0)e +
\shalf(1+\dot{x}_0)e^{1/2}
               \g_\pp + \shalf(1-\dot{x}_0)e^{1/2} \g_\mm \non \\
        &~& + \shalf(1+\dot{x}_0) {{\tilde{\psi}}_\pp}^+\g_+
      -\shalf(1-\dot{x}_0) {{\tilde{\psi}}_\mm}^-\g_-   + \frac{g}{4}] \non \\
I^{(1)} &=& -m \int d\tau [-\sihalf g^{-1}(1+\dot{x}_0)(1-\dot{x}_0)e^{3/4}
                \{ {{\tilde{\psi}}_\pp}^+ \th_0^+ - {{\tilde{\psi}}_\mm}^-
\th_0^- \} \non \\
         &~&  +\sihalf(1+\dot{x}_0)
                  {{\tilde{\psi}}_\pp}^+ \th_0^- e^{1/4}- \sihalf(1-\dot{x}_0)
                  {{\tilde{\psi}}_\mm}^- \th_0^+ e^{1/4} ] \non \\
I^{(2)} &=& -m \int d\tau [ -ig^{-1}e\{ (1-\dot{x}_0) ({\cal{D}}\th_0^+)
\th_0^+
             + (1-\dot{x}_0) ({\cal{D}}\th_0^-)\th_0^-\}
              +i e^{1/2}\{ ({\cal{D}}\th_0^+)\th_0^- +
({\cal{D}}\th_0^-)\th_0^+ \} \non \\
         &~&     +\sihalf g^{-1} e \tilde{A}(1+\dot{x}_0)(1-\dot{x}_0)\th_0^+
\th_0^-
               -2 g^{-1} e^{1/2}(1+\dot{x}_0)(1-\dot{x}_0)
{{\tilde{\psi}}_\pp}^+
                {{\tilde{\psi}}_\mm}^- \th_0^+ \th_0^- ] \non \\
I^{(3)} &=& -m \int d\tau [4 g^{-1}e^{3/4}\{ (1-\dot{x}_0)
                  {{\tilde{\psi}}_\mm}^- ({\cal{D}}\th_0^+) + (1+\dot{x}_0)
                  {{\tilde{\psi}}_\pp}^+ ({\cal{D}}\th_0^-) \} \th_0^+ \th_0^-
] \non \\
I^{(4)} &=& -m \int d\tau [ 16g^{-1}e({\cal{D}}\th_0^+) ({\cal{D}}\th_0^-)
                   \th_0^+ \th_0^-]
\eea
It is simple to obtain the component form of the dilaton supergravity part of
the
action in conformal gauge from \reff{dil} using standard techniques
\cite{bible}.

\sect{\bf Component Results and Non-Triviality of Solution}

Ordinarily, to go from superspace results to
 the usual ${x}$-space components, one uses the standard technique of
 fixing a Wess-Zumino gauge ($\Del_\pm| = \di_\pm$).  However, as is discussed
 in \cite{rocek}, the usual procedure needs to be modified somewhat in order
to find a Wess-Zumino gauge choice such that the superconformal
gauge ($E_\pm = e^S D_\pm$) is compatible with the ordinary $x$-space conformal
gauge
($e_a^m = \rho \delta_a^m$).  Using these results, we can write $E_\a$ in
component
conformal gauge as
\bea
E_\pm &=& e^S D_\pm \non \\
 &=& [ e^{-1/4} + \sihalf(\th^+ {\psi_\pp}^+ - \th^- {\psi_\mm}^-) +
\squart \th^+ \th^-
            e^{1/4}(iA + {\psi_\mm}^- {\psi_\pp}^+)] D_\pm
\eea
 Taylor expanding $e^S$ allows us to
 identify the gravitino (${\psi_\pp}^+, {\psi_\mm}^-$) and the
 component auxiliary field of the supergravity multiplet, $A$. Setting $\kappa
= 8$, we find
\bea
e^{-1/4} &=& e^{-M|x-x_0|} \non \\
\shalf {\psi_\pp}^+ &=&  e^{-M|x-x_0|}M \th_0^+ \e(x-x_0) \non \\
\shalf {\psi_\mm}^- &=&  e^{-M|x-x_0|}M \th_0^- \e(x-x_0)  \\
\squart e^{1/4}i A  &=&   2e^{-M|x-x_0|}M \th_0^+ \th_0^- \d(x-x_0) \non \\
\squart e^{1/4} {\psi_\mm}^- {\psi_\pp}^+  &=& - e^{-M|x-x_0|}M^2 \th_0^+
\th_0^- \non \label{components}
\eea

   We can now compute
 the scalar supercurvature, $R$, of the
resulting background. From \reff{R}, we find
\beq
R = 8M \d(x-x_0) \d^{(2)}(\th-\th_0)
\eeq
which can be rewritten using \reff{components} as
\beq
R = e^{1/4}[ e^{1/4}iA + 2 \th^-\di {\psi_\pp}^+ - 2 \th^+\di {\psi_\mm}^-
     -4 \th^+ \th^-(\di^2 e^{-1/4} - e^{-1/4}M^2)]
\eeq
and compared with the general results in \cite{brooks, howe}.
The $x$-space Ricci curvature, $r$, is contained in $\Del^2 R|$ .
We find
\bea
\Del^2 R | &=& 4 e^{-1/4}(\di^2 e^{-1/4} - e^{-1/4}M^2) \non \\
           &=& - 8M \d(x-x_0) \non \\
           &=& r
\eea
since the other terms in $\Del^2 R|$, involving the gamma-trace of the
gravitino, the
curl of the gravitino and the auxiliary field, vanish
in this simple case.

 We turn now to the non-triviality of the solution -- specifically, the fact
that the
solution  for $S$ cannot be obtained by an infinitesimal supersymmetry
transformation from the
bosonic one.
 This can be seen in one of two ways: from a purely superspace argument or from
components. In the former case, we note that the
superspace
described by $S$ has torsion, as expected in a {\it bona fide} supergravity
solution. As the torsion ${T_{AB}}^C$ is a supercovariant quantity,  its value
remains unchanged under a suitable gauge transformation. Such a gauge
transformation
might set the gravitino to zero, but then simultaneously there must be a
redefinition of the
vielbein so that overall ${T_{AB}}^C$ is unchanged.

  Alternatively, a supergravity solution can be seen to be trivial if one can
find an infinitesimal
spinor $\a(x)$ such that $\psi_a^\mu = D_a\a^\mu$ where $D_a = e_a^m \di_m +
\o_a M$
is the component covariant gravitational derivative \cite{aichel}.
 For example, the differential equation
 that $\a^+$ must satisfy,
${\psi_\pp}^+ = D_\pp \a^+$, becomes, using the above results
\beq
\di[e^{-M|x-x_0|}\a^+] = 2M \th_0^+ \e(x-x_0)
\eeq
with the solution $\a^+ = 2M\th_0^+ e^{M|x-x_0|} |x-x_0|$. We note that $\a^+$
is not well
behaved at infinity, and so the solution for $S$ is not related to a trivial
one
by an infinitesimal supersymmetry transformation.

\sect{Computation of Vielbein}

   In each of the cases considered in section 8, the vielbein takes the form
\bea
{{\calE'}_M}^A = \left[
\begin{array}{cccc}
(e^{-2S})' & 0 & -i(D_+e^{-S})' & -i(D_+e^{-S})' \\
0 & {{\calE'}_u}^u & {{\calE'}_u}^+ & {{\calE'}_u}^- \\
 {{\calE'}_+}^t  &  {{\calE'}_+}^u & {{\calE'}_+}^+  &  {{\calE'}_+}^- \\
 {{\calE'}_-}^t   & {{\calE'}_-}^u   & {{\calE'}_-}^+ & {{\calE'}_-}^-
\end{array}
\right]
\eea
where $(X)'$ means the object is to be evaluated in the transformed
coordinates, and the
rows and columns of the vielbein are labelled by $(t, u, +, -)$.  We
note that in
obtaining ${{\calE'}_M}^A$ in its final form, we have performed two basis
changes
(one world,
one tangent) in addition to the actual supercoordinate transformation.  We
reiterate that
$\e(x-x_0) = \e(u-u_0)$ by construction, and that $\d(x-x_0) = |C|\d(u-u_0)$.

 For $C>0$, we find
\bea
(e^{-2S})' &=& \frac{2m}{|C|} |u-u_0 - i(\eta^+ \eta_0^+ - \eta^- \eta_0^-)| +
1 \non \\
(e^{-S})' &=& \sqrt{\frac{2m}{|C|} |u-u_0 - i(\eta^+ \eta_0^+ - \eta^-
\eta_0^-)| + 1} \non \\
(D_+e^{-S})' &=& -(e^{-S})'(D_+S)' \non \\
(D_-e^{-S})' &=& -(e^{-S})'(D_-S)' \\
(D_+S)'&=& -2M(\th^+ - \th_0^+)[ i\e(u-u_0) - 2\th^-\th_0^- |C|\d(u-u_0)] \non
\\
(D_-S)'&=& 2M(\th^- - \th_0^-)[ i\e(u-u_0) - 2\th^+\th_0^+ |C|\d(u-u_0)] \non
\eea
where
\bea
|x -x_0| &=& \frac{1}{2m} ln(\frac{2m}{|C|}|u-u_0| + 1) \non \\
\th^\pm &=& \frac{\sqrt{|C|}}{2m}\frac{\eta^\pm}{\left[|u-u_0| + |C|/2m \pm
\sihalf
      \eta^\mp \eta_0^\mp \e(u-u_0) \right]} \non\\
\th_0^\pm &=& \frac{\eta^\pm_0}{\sqrt{|C|}}
\eea
In addition, we also have
\bea
{{\calE'}_u}^u &=& \frac{1}{2m} \frac{\e(u-u_0)}{|u-u_0| + |C|/2m}(e^{-2S})'
\non \\
{{\calE'}_u}^+ &=& \frac{1}{2m} \frac{\e(u-u_0)}{|u-u_0| +
|C|/2m}(-iD_+e^{-S})'
               -\frac{\sqrt{|C|}}{2m}\frac{\eta^+}{[X]^2}(\e(u-u_0) +i
\eta^-\eta_0^-\d(u-u_0))
                 (e^{-S})' \non \\
{{\calE'}_u}^- &=& \frac{1}{2m} \frac{\e(u-u_0)}{|u-u_0| + |C|/2m}(iD_-e^{-S})'
               -\frac{\sqrt{|C|}}{2m}\frac{\eta^-}{[Y]^2}(\e(u-u_0) -i
\eta^+\eta_0^+\d(u-u_0))
                        (e^{-S})' \non \\
{{\calE'}_+}^t &=&  \frac{\sqrt{|C|}}{2m} \frac{1}{[X]}(-ie^{-2S} \th^+)' \non
\\
{{\calE'}_+}^u &=&  \frac{\sqrt{|C|}}{2m} \frac{1}{[X]}(-ie^{-2S} \th^+)'  \non
\\
{{\calE'}_+}^+ &=&  \frac{\sqrt{|C|}}{2m} \frac{1}{[X]}(e^{-S})'(1
-2(D_+S))\th^+)' \non \\
{{\calE'}_+}^- &=&  \frac{\sqrt{|C|}}{2m} \frac{\eta^-}{[Y]^2}(-\sihalf
\eta_0^+
\e(u-u_0))(e^{-S})'
                 \non\\
{{\calE'}_-}^t &=&  \frac{\sqrt{|C|}}{2m} \frac{1}{[Y]}(-ie^{-2S} \th^-)'\non
\\
{{\calE'}_-}^u &=& - \frac{\sqrt{|C|}}{2m} \frac{1}{[Y]}(-ie^{-2S} \th^-)'\non
\\
{{\calE'}_-}^+ &=&  \frac{\sqrt{|C|}}{2m} \frac{\eta^+}{[X]^2}(\sihalf \eta_0^-
\e(u-u_0))(e^{-S})'
                 \non\\
{{\calE'}_-}^- &=&  \frac{\sqrt{|C|}}{2m} \frac{1}{[Y]}(e^{-S})'(1
-2(D_-S))\th^-)'
\eea
where we denote
\bea
X &=& |u-u_0| + |C|/2m + \sihalf \eta^- \eta_0^- \e(u-u_0) \non \\
Y &=& |u-u_0| + |C|/2m - \sihalf \eta^+ \eta_0^+ \e(u-u_0)
\eea

 For $C<0$, we have the three different regions:

{\bf Region (i)}:
\bea
(e^{-2S})' &=& \frac{-2m}{|C|} [u-u_0 - i(\eta^+ \eta_0^+ - \eta^- \eta_0^-)] -
1 \non \\
(e^{-S})' &=& \sqrt{\frac{-2m}{|C|} [u-u_0 - i(\eta^+ \eta_0^+ - \eta^-
\eta_0^-)] - 1} \non \\
(D_+e^{-S})' &=& -(e^{-S})'(D_+S)' \non \\
(D_-e^{-S})' &=& -(e^{-S})'(D_-S)' \\
(D_+S)'&=& -2M(\th^+ - \th_0^+)[ i\e(u-u_0) - 2\th^-\th_0^- |C|\d(u-u_0)] \non
\\
(D_-S)'&=& 2M(\th^- - \th_0^-)[ i\e(u-u_0) - 2\th^+\th_0^+ |C|\d(u-u_0)] \non
\eea
where
\bea
x -x_0 &=& \frac{-1}{2m} ln(\frac{-2m}{|C|}(u-u_0) - 1) \non \\
\th^\pm &=& \frac{\sqrt{|C|}}{2m}\frac{i\eta^\pm}{\left[(u-u_0) + |C|/2m \pm
\sihalf
           \eta^\mp \eta_0^\mp \right]} \non\\
\th_0^\pm &=& \frac{i\eta^\pm_0}{\sqrt{|C|}}
\eea
In addition, we also have
\bea
{{\calE'}_u}^u &=& \frac{-1}{2m} \frac{1}{(u-u_0) + |C|/2m}(e^{-2S})'  \non \\
{{\calE'}_u}^+ &=& \frac{-1}{2m} \frac{1}{(u-u_0) + |C|/2m}(-iD_+e^{-S})'
               -\frac{\sqrt{|C|}}{2m}\frac{i\eta^+}{[X]^2}
                 (e^{-S})' \non \\
{{\calE'}_u}^- &=& \frac{-1}{2m} \frac{1}{(u-u_0) + |C|/2m}(iD_-e^{-S})'
               -\frac{\sqrt{|C|}}{2m}\frac{i\eta^-}{[Y]^2}
                        (e^{-S})' \non \\
{{\calE'}_+}^t &=&  \frac{\sqrt{|C|}}{2m} \frac{i}{[X]}(-ie^{-2S}\th^+)' \non
\\
{{\calE'}_+}^u &=&  \frac{\sqrt{|C|}}{2m} \frac{i}{[X]}(-ie^{-2S}\th^+)' \non
\\
{{\calE'}_+}^+ &=&  \frac{\sqrt{|C|}}{2m} \frac{i}{[X]}(e^{-S})'(
1-2(D_+S)\th^+)' \non \\
{{\calE'}_+}^- &=&  \frac{\sqrt{|C|}}{2m} \frac{i\eta^-}{[Y]^2}(-\sihalf
\eta_0^+)(e^{-S})'
                 \non\\
{{\calE'}_-}^t &=&  \frac{\sqrt{|C|}}{2m} \frac{i}{[Y]}(-ie^{-2S}\th^-)' \non
\\
{{\calE'}_-}^u &=& - \frac{\sqrt{|C|}}{2m} \frac{i}{[Y]}(-ie^{-2S}\th^-)' \non
\\
{{\calE'}_-}^+ &=&  \frac{\sqrt{|C|}}{2m} \frac{i\eta^+}{[X]^2}(\sihalf
\eta_0^-)(e^{-S})'
                 \non\\
{{\calE'}_-}^- &=&  \frac{\sqrt{|C|}}{2m} \frac{i}{[Y]}(e^{-S})' (
1-2(D_-S)\th^-)'
\eea
where we denote
\bea
X &=& (u-u_0) + |C|/2m + \sihalf \eta^- \eta_0^-  \non \\
Y &=& (u-u_0) + |C|/2m - \sihalf \eta^+ \eta_0^+
\eea

{\bf Region (ii)}:

\bea
(e^{-2S})' &=& \frac{-2m}{|C|} |u-u_0 - i(\eta^+ \eta_0^+ - \eta^- \eta_0^-)| +
1 \non \\
(e^{-S})' &=& \sqrt{\frac{-2m}{|C|} |u-u_0 - i(\eta^+ \eta_0^+ - \eta^-
\eta_0^-)| + 1} \non \\
(D_+e^{-S})' &=& -(e^{-S})'(D_+S)' \non \\
(D_-e^{-S})' &=& -(e^{-S})'(D_-S)' \\
(D_+S)'&=& -2M(\th^+ - \th_0^+)[ i\e(u-u_0) - 2\th^-\th_0^- |C|\d(u-u_0)] \non
\\
(D_-S)'&=& 2M(\th^- - \th_0^-)[ i\e(u-u_0) - 2\th^+\th_0^+ |C|\d(u-u_0)] \non
\eea
where
\bea
x -x_0 &=& \frac{-1}{2m} ln(1 -\frac{2m}{|C|}|u-u_0|) \non \\
\th^\pm &=& \frac{\sqrt{|C|}}{2m}\frac{\eta^\pm}{\left[|u-u_0| - |C|/2m \pm
\sihalf
      \eta^\mp \eta_0^\mp \e(u-u_0) \right]} \non\\
\th_0^\pm &=& \frac{-\eta^\pm_0}{\sqrt{|C|}}
\eea
In addition, we also have
\bea
{{\calE'}_u}^u &=& \frac{-1}{2m} \frac{\e(u-u_0)}{|u-u_0| - |C|/2m}(e^{-2S})'
\non \\
{{\calE'}_u}^+ &=& \frac{-1}{2m} \frac{\e(u-u_0)}{|u-u_0| -
|C|/2m}(-iD_+e^{-S})'
               -\frac{\sqrt{|C|}}{2m}\frac{\eta^+}{[X]^2}(\e(u-u_0) +i
\eta^-\eta_0^-\d(u-u_0))
                 (e^{-S})' \non \\
{{\calE'}_u}^- &=& \frac{-1}{2m} \frac{\e(u-u_0)}{|u-u_0| -
|C|/2m}(iD_-e^{-S})'
               -\frac{\sqrt{|C|}}{2m}\frac{\eta^-}{[Y]^2}(\e(u-u_0) -i
\eta^+\eta_0^+\d(u-u_0))
                        (e^{-S})' \non \\
{{\calE'}_+}^t &=&  \frac{\sqrt{|C|}}{2m} \frac{1}{[X]}(-ie^{-2S}\th^+)' \non
\\
{{\calE'}_+}^u &=&  \frac{\sqrt{|C|}}{2m} \frac{1}{[X]}(-ie^{-2S}\th^+)'  \non
\\
{{\calE'}_+}^+ &=&  \frac{\sqrt{|C|}}{2m} \frac{1}{[X]}(e^{-S})'(1-2(D_+S)
\th^+)' \non \\
{{\calE'}_+}^- &=&  \frac{\sqrt{|C|}}{2m} \frac{\eta^-}{[Y]^2}(-\sihalf
\eta_0^+
\e(u-u_0))(e^{-S})'
                 \non\\
{{\calE'}_-}^t &=&  \frac{\sqrt{|C|}}{2m} \frac{1}{[Y]}(-ie^{-2S}\th^-)' \non
\\
{{\calE'}_-}^u &=& - \frac{\sqrt{|C|}}{2m} \frac{1}{[Y]}(-ie^{-2S}\th^-)' \non
\\
{{\calE'}_-}^+ &=&  \frac{\sqrt{|C|}}{2m} \frac{\eta^+}{[X]^2}(\sihalf \eta_0^-
\e(u-u_0))(e^{-S})'
                 \non\\
{{\calE'}_-}^- &=&  \frac{\sqrt{|C|}}{2m} \frac{1}{[Y]}(e^{-S})'(1-2(D_-S)
\th^-)'
\eea
where we denote
\bea
X &=& |u-u_0| - |C|/2m + \sihalf \eta^- \eta_0^- \e(u-u_0) \non \\
Y &=& |u-u_0| - |C|/2m - \sihalf \eta^+ \eta_0^+ \e(u-u_0)
\eea

{\bf Region (iii)}:
\bea
(e^{-2S})' &=& \frac{2m}{|C|} [u-u_0 - i(\eta^+ \eta_0^+ - \eta^- \eta_0^-)] -
1 \non \\
(e^{-S})' &=& \sqrt{\frac{2m}{|C|} [u-u_0 - i(\eta^+ \eta_0^+ - \eta^-
\eta_0^-)] - 1} \non \\
(D_+e^{-S})' &=& -(e^{-S})'(D_+S)' \non \\
(D_-e^{-S})' &=& -(e^{-S})'(D_-S)' \\
(D_+S)'&=& -2M(\th^+ - \th_0^+)[ i\e(u-u_0) - 2\th^-\th_0^- |C|\d(u-u_0)] \non
\\
(D_-S)'&=& 2M(\th^- - \th_0^-)[ i\e(u-u_0) - 2\th^+\th_0^+ |C|\d(u-u_0)] \non
\eea
where
\bea
x -x_0 &=& \frac{1}{2m} ln(\frac{2m}{|C|}(u-u_0) - 1) \non \\
\th^\pm &=& \frac{\sqrt{|C|}}{2m}\frac{i\eta^\pm}{\left[(u-u_0) - |C|/2m \pm
\sihalf
           \eta^\mp \eta_0^\mp \right]} \non\\
\th_0^\pm &=& \frac{-i\eta^\pm_0}{\sqrt{|C|}}
\eea
In addition, we also have
\bea
{{\calE'}_u}^u &=& \frac{1}{2m} \frac{1}{(u-u_0) - |C|/2m}(e^{-2S})' \non \\
{{\calE'}_u}^+ &=& \frac{1}{2m} \frac{1}{(u-u_0) - |C|/2m}(-iD_+e^{-S})'
               -\frac{\sqrt{|C|}}{2m}\frac{i\eta^+}{[X]^2}
                 (e^{-S})' \non \\
{{\calE'}_u}^- &=& \frac{1}{2m} \frac{1}{(u-u_0) - |C|/2m}(iD_-e^{-S})'
               -\frac{\sqrt{|C|}}{2m}\frac{i\eta^-}{[Y]^2}
                        (e^{-S})' \non \\
{{\calE'}_+}^t &=&  \frac{\sqrt{|C|}}{2m} \frac{i}{[X]}(-ie^{-2S}\th^+)' \non
\\
{{\calE'}_+}^u &=&  \frac{\sqrt{|C|}}{2m} \frac{i}{[X]}(-ie^{-2S}\th^+)' \non
\\
{{\calE'}_+}^+ &=&  \frac{\sqrt{|C|}}{2m}
\frac{i}{[X]}(e^{-S})'(1-2(D_+S)\th^+)' \non \\
{{\calE'}_+}^- &=&  \frac{\sqrt{|C|}}{2m} \frac{i\eta^-}{[Y]^2}(-\sihalf
\eta_0^+)(e^{-S})'
                 \non\\
{{\calE'}_-}^t &=&  \frac{\sqrt{|C|}}{2m} \frac{i}{[Y]}(-ie^{-2S}\th^-)' \non
\\
{{\calE'}_-}^u &=& - \frac{\sqrt{|C|}}{2m} \frac{i}{[Y]}(-ie^{-2S}\th^-)' \non
\\
{{\calE'}_-}^+ &=&  \frac{\sqrt{|C|}}{2m} \frac{i\eta^+}{[X]^2}(\sihalf
\eta_0^-)(e^{-S})'
                 \non\\
{{\calE'}_-}^- &=&  \frac{\sqrt{|C|}}{2m}
\frac{i}{[Y]}(e^{-S})'(1-2(D_-S)\th^-)'
\eea
where we denote
\bea
X &=& (u-u_0) - |C|/2m + \sihalf \eta^- \eta_0^-  \non \\
Y &=& (u-u_0) - |C|/2m - \sihalf \eta^+ \eta_0^+
\eea

\end{document}